# Role of Zn in obtaining of semi-insulating CdZnTe crystals for ionizing radiations detectors

O.A. Matveev, V.E. Sedov, V.P. Karpenko, A.I. Terent'ev, N.K. Zelenina, A.A.Tomasov

A.F.Ioffe Physico-Technical Institute RAS, Politecknicheskaya, 26, 194021,St-Petersburg, Russia

Study of a photoluminescence and the extrinsic photoconductivity has shown that excess concentration of $V_{Cd}$ in n-CdZnTe crystals is the main reason of low value of $\mu_h\tau_h$. Reduction of concentration of $V_{Cd}$ (decreasing of intensity of ~1eV photoluminescence band and intensity of the extrinsic photoconductivity (0.9-1.3) eV band) by annealing of the crystals at 600°C results in increasing of value of $\mu_h\tau_h$. Influence of Zn on formation of the basic photoelectric properties of crystals CdZnTe has been explained by "self-control" of a concentration of cadmium vacancies that makes process of self-compensation less dependent from pressure $P_{Cd}$ in comparison with CdTe crystals.

Recently, major interest has appeared to crystals of $Cd_{1-x}Zn_xTe$ solid solutions, which have a wider than CdTe band gap that allows increasing a value of an electric field in detectors considerably and essentially improving performance of the devices. However obtained solid solution crystals have much smaller [1] values of $\mu_h\tau_h$ than CdTe that essentially limits a range of application of CdZnTe detectors. In this connection there is actual a question of finding - out of the reasons resulting in low value of $\mu_h\tau_h$ in CdZnTe crystals.

The present work is devoted to studying of photoelectric and luminescent properties of CdTe and $Cd_{1-x}Zn_xTe$ crystals with $x \leq 0.1$, Cl doped ($2\times10^{16}$-$10^{17}$ cm$^{-3}$) and undoped. Samples for examinations were prepared from the crystals, which have been grown by a method of a horizontal directed crystallization under controllable pressure of cadmium vapors [2]. The previously tested (Hall effect, photoluminescence, photoconductivity, time-of-flight technique) samples then were annealed at 600°C in isothermal conditions during 3 hours with the subsequent quenching or slow cooling up to a room temperature during 48 hours. Annealing was carried out in an evacuated quartz ampoule with filling from crushed crystals from the same ingot to avoid evaporation of a material from a surface of the sample and to maintain a composition of defects maximally.

All studied CdTe:Cl and $Cd_{1-x}Zn_xTe$:Cl crystals were semi-insulating n-or p-type conductance crystals with concentration of charge carriers of $10^7$-$10^9$ cm$^{-3}$, and undoped crystals were p-type with concentration of $10^{15}$ cm$^{-3}$ before annealing. After annealing with quenching all samples became low resistivity n-type conductance crystals. Annealing with slow cooling has again transferred almost all crystals except of crystals not doped with chlorine into a semi-insulating condition.

It is possible to explain these results as follows. In an initial after growth state of crystal donors are known [3] to form associates with acceptors (vacancy of metal) and crystals become semi-insulating with a high degree of compensation. The heating temperature 600°C is enough temperature for these associates were destroyed and such a state is maintained after quenching. The crystal becomes low resistivity n-type conductance one since donors (Cl) create shallow levels, which are completely ionized, and acceptors (vacancy of metal) are enough deep centers. However, if annealed at 600°C and quenched crystal is heat again and then slow cooled the defects will interact forming associates and the crystal becomes semi-insulating again..

An influence of Zn on performance of analog detectors (kinetics of photocurrent

Corresponding author: e-mail: Oleg.Matveev@mail.ioffe.ru Phone: +7 812 247 91 89, Fax: +7 812 247 10 17

pulse and value of photocurrent memory) was experimentally shown by us [4]. However, the concrete mechanism of Zn action is not clear. Zn in CdTe is an isoelectronic impurity replacing Cd. Isoelectronic impurities may be traps for nonequilibrium holes or electrons if there is simultaneous significant distinction of two parameters of the basic and replacing atom: electron affinities and covalent radius [5]. Covalent radiuses of Cd and Zn differ really strongly (1.49 and 1.28 A°, correspondingly). However values of electron affinity for them practically coincide (1.69 for Cd and 1.65 for Zn). Thus, introduction of Zn in CdTe in this case should not result in appearance of any levels in the band gap. But another mechanism of isoelectronic impurities action is also possible. Discrepancy of covalent radiuses of the basic and replacing atom causes elastic strains in a lattice of the crystal and change of native defects concentration. Thus, action of Zn on structure of energetic levels in the band gap of the crystal may occur as a result of some specific influence of Zn on processes of formation of native defects that facilitates the process of self-compensation. This feature of Zn activity in solid solutions CdZnTe was found in [6] by learning of positrons annihilation. Even slight addition of Zn in CdTe crystal was shown to lead to origination of neutral divacancies of metal. Their concentration increases in accordance with the additive Zn concentration. In the same work it was shown that there are only monovacancies of metal, which form associations with donors in crystals CdTe not doped with Zn. The mechanism of integrating of monovacancies into divacancies is unknown. It is probably result of interaction of vacancies with a field of the elastic stresses caused by a large mismatch of covalent radiuses of Cd and Zn.

Naturally, that there is equilibrium divacancy -vacancy in crystals CdZnTe and some part of neutral divacancies dissociates on monovacancies of metal which will interact with donors or remain in a free state. These divacancies of metal serve as a reservoir from which the necessary quantity of monovacancies arises in a crystal in accordance with self-compensation conditions.

Divacancies of cadmium are not formed in CdTe. The concentration of vacancies of metal in CdTe is lower in comparison with CdZnTe (ZnTe) [7]. Therefore the high degree of association of native defects (self-compensation) with formation of neutral complexes takes place in CdTe in a narrower interval of cadmium pressures in comparison with CdZnTe. Hence, during obtaining of CdTe crystals cadmium pressure should be strictly determined both at growing of a crystal and at its annealing. Otherwise the crystal will be low-resistivity n-or p-type conductance one.

Monovacancy of Cd is the double-charged acceptor, which can be in two charge states: $V_{Cd}^{-1}$ and $V_{Cd}^{-2}$ depending on a Fermi level position. Two levels in band gap Ev+0.06 eV and Ec-0.6 eV [8-11] correspond to these states. The center $V_{Cd}^{-2}$ should actively capture holes but the subsequent electron capture is complicated because of a Coulomb repulsion by an electron remaining on the center. The behavior of a similar double acceptor was described in [11]. Such centers reduce a lifetime of holes and may result in occurrence of photocurrent memory.

Above mentioned allows understanding why at crystal growth of CdZnTe solid solutions at high pressure of inert gas [12-13] practically in absence of control of pressure of a volatile component (Cd), semi-insulated crystals of detector grade with high $\mu_e\tau_e$ values are obtained. From the same reasons it is clear why these crystals have low values of $\mu_h\tau_h$ as a rule. Really dissociation of neutral divacancies of metal should result in increasing of concentration of metal monovacancies [6], which sharply reduce a lifetime of holes especially in case of uncontrolled pressure of cadmium.

Our study of CdTe and CdZnTe crystals revealed extrinsic photoconductivity in the spectral range of (0.6-1.3) eV. We believe that this photoconductivity band is related with the presence of metal vacancy ($V_{Cd}^{-2}$). As it is seen from the photoconductivity spectra (Fig.1) the intensity of this band increases with increasing of Zn content. Be-

*

sides this, it may be seen that annealing of crystals with low Zn concentration (x=2 ×10$^{-4}$) resulted in decreasing of intensity of this band. In detectors produced from these crystals the photocurrent memory diminished (down to ~0.1%), the shape of photo response pulse is improved, the value of $\mu\tau_h$ increased up to ~3× 10$^{-5}$cm$^2$/V (it was ~1× 10$^{-5}$ cm$^2$/V before annealing).

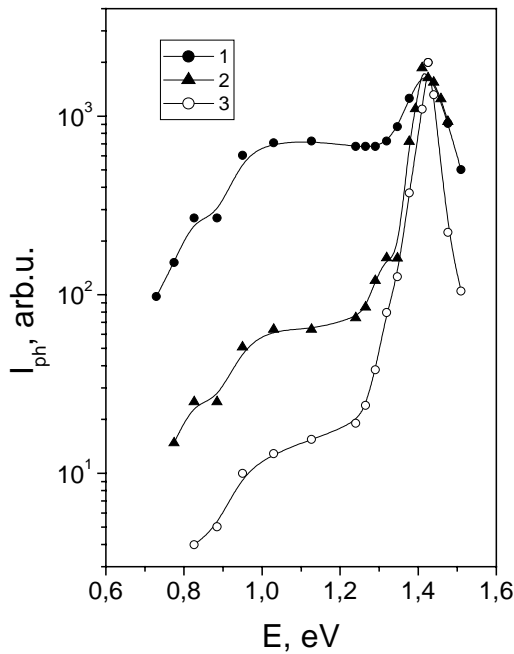

Fig.1. Spectra of extrinsic photoconductivity of $Cd_{1-x}Zn_xTe:Cl$ crystals for tomographic detectors; curve 1 – x=0.1; curve 2 – x=0.0002, curve 3 – after annealing, x=0.0002.

The PL spectrum for all samples consisted of 3 bands (Table 1): I – edge emission with the peak $\hbar\omega_m \approx 1.6$ eV, II – the most intensive band with $\hbar\omega_m \approx 1.45$ eV and III – the band with $\hbar\omega_m \approx 1.0$ eV.

**Table 1**. PL parameters of samples from different groups of crystals: CdTe:Cl, CdZnTe:Cl,.

| Samples | PL band | Position of maximum, $\hbar\omega_m$, eV | FWHM, $\hbar\omega$, meV | Arbitrary intensity of band, % |
|---|---|---|---|---|
| n-CdTe:Cl N° 250 | I | 1,58 | --- | 26 |
| | II | 1,427 | 100 | 100 |
| | III | 1,057 | 228 | 2,0 |
| n-CdZnTe:Cl, 10$^{-4}$ % Zn, N° 337 | I | 1,58 | --- | 31 |
| | II | 1,434 | 96 | 100 |
| | III | 1,088 | 216 | 12,8 |
| n-CdZnTe:Cl, 5 % Zn, N° 340 | I | 1,59 | --- | 18 |
| | II | 1.438 | 100 | 100 |
| | III | 1,060 | 190 | 3,8 |
| n-CdZnTe:Cl 10 %Zn, N° 319 | I | 1,63 | --- | 58,3 |
| | II | 1,470 | 110 | 100 |
| | III | 1,068 | 157 | 24,8 |

The Table 1 suggests as follows.
 1.The peak of the edge PL band I has the same position for all samples. Doping with Zn shifts it to higher energies in accordance with increasing of the band gap. The relative intensity of the edge emission as well as the integral PL intensity is the largest for the samples with 10% of Zn. This suggests that these samples have the lowest con-

*

centration of nonradiative recombination centers.

2. The band II has a trend of shifting to short–wave range with increasing of Zn content too. As it is known [3] this center has the energy level $E_V+0.15eV$ as is a trap for holes. The peak position spreading for II band may be related with formation of complex agglomeration including more than one $V_{Cd}$. In [14] it is supposed that two different centers of the nature unknown up to now contribute to band II. They may be agglomeration with $V_{Cd}$ and $V_{Te}$.

3. Band III apparently does not shift with Zn doping. There is a large spreading of its intensity and peak position for different samples. The PL band with the peak at ~1eV may be related with the transition of holes from the valence band onto the $E_C-0.6$ eV level, which is usually attributed to double acceptor $V_{Cd}^{-2}$ [10,11]. Probably this center appears also in photoconductivity spectra (Fig.1). It is not clear up to now how rather high concentration of such centers (intensive III band – see the Table 1) in the samples with Cl and Zn is compatible with their good photoelectric properties. However, appreciable improvement of photoelectric properties of the sample N°337 as a result of annealing with quenching following by annealing with slow cooling was accompanied by dramatic (more than order in magnitude) reducing of the III band intensity. It is noteworthy the ~40 meV shift of 1.45 eV band to lower energies as a result of such thermo treatment. Probably, it may be caused by some changes in the constitution of the centers responsible for the 1.45 eV band. In any case, the concentration of both the trapping centers and the centers of recombination diminished after the thermo treatment mentioned above.

Results of the present work have allowed understanding in basically the mechanism of influence of Zn on formation of the main photoelectrical properties of semi-insulating CdZnTe crystals and detectors on their basis.

Addition of Zn results in formation of divacancies of metal, which in part dissociate and provide a crystal with necessary quantity of vacancies for processes of complex formation. It facilitates process of self-compensation and makes it less dependent from $P_{Cd}$. Thus apparently more full association of defects occurs. However, if $P_{Cd}$ is not controlled as at growth of crystals by the HPB method an exuberant concentration of $V_{Cd}$ (intensive PL band ~ 1 eV) may be created in the crystal that will result in low value of $\mu_h\tau_h$. And if the semi-insulating crystal is p-type conductivity $V_{Cd}$ becomes the recombination center and it will reduce a value of $\mu_e\tau_e$.

Thus Zn directly not creating energetic levels rather strongly allows to influence process of self-compensation and complex formation and so to influence formation of electronic properties of crystals. However to achieve the high values of $\mu_h\tau_h$ and $\mu_e\tau_e$ in crystals, low size of memory of a photocurrent and high sensitivity of detectors it is necessary to reduce concentration of free vacancies of metal as annealing experiments with CdZnTe crystals have shown.

**Acknowledgements.** The authors greatly appreciate financial support of this work from INTAS Grant #99-01456.**Acknowledgements.** The authors greatly appreciate financial support of this work from INTAS Grant #99-01456.

*

*